\documentclass[aps,prb,twocolumn,superscriptaddress]{revtex4-1}  
\usepackage{graphicx}  
\usepackage{dcolumn}   
\usepackage{bm}        
\usepackage{amssymb}   
\usepackage{hyperref}
\usepackage{physics}
\usepackage{mathtools}

\newcommand{\perptm}{\prescript{}{\perp}{\mathrm{TM}}}
\newcommand{\paralleltm}{\prescript{}{\parallel}{\mathrm{TM}}}
\newcommand{\tmat}[2]{\prescript{}{\mathrm{#1}}{\mathbf{#2}}}

\usepackage{xcolor}

\begin{document}


\title{Resolving the two-dimensional ANNNI model using transfer matrices}

\author{Yi Hu}
\affiliation{Department of Chemistry, Duke University, Durham, North Carolina 27708, USA}
\author{Patrick Charbonneau}
\email{patrick.charbonneau@duke.edu}
\affiliation{Department of Chemistry, Duke University, Durham, North Carolina 27708, USA}
\affiliation{Department of Physics, Duke University, Durham, North Carolina 27708, USA}
\date{\today}

\begin{abstract}
The phase diagram of the two-dimensional ANNNI model has long been theoretically debated. Extremely long structural correlations and relaxation times further result in numerical simulations making contradictory predictions. Here, we introduce a numerical transfer matrix treatment that bypasses these problems, and thus overcome various ambiguities in the phase diagram. In particular, we confirm the transition temperatures and clarify the order of the transition to the floating incommensurate phase. Our approach motivates considering transfer matrices for resolving long-standing problems in related statistical physics models.
\end{abstract}

\maketitle

\section{Introduction}

Patterned and modulated phases robustly form when the components of a system interact via competing short-range attractive and long-range repulsive (SALR) interactions.~\cite{sciortino2004equilibrium,ciach2013origin,zhuang2016recent,royall2018hunting}
Such phases have indeed been observed in materials ranging from magnetic alloys,~\cite{seul1995domain,portmann2003inverse} to lipidic surfactants,~\cite{caffrey2009crystallizing,fink2019structure} and biological tissues.~\cite{lecuit2007cell,heisenberg2013forces} Frustration, however, is also associated with slowly decaying finite-size corrections and to complex relaxation processes, which both severely impede the study of equilibrium phases by numerical simulations. Specialized sampling techniques are thus needed to study even minimal microphase-formers,~\cite{zhang2010monte,shirakura2014kosterlitz,zhuang2016equilibriumPRL,lei2019event} let along more realistic ones.

Lattice models with SALR interactions were first formulated forty years ago, the simplest being the axial next-nearest-neighbor Ising (ANNNI) model.~\cite{fisher1980infinitely} 
Yet even in two dimensions, this small perturbation to the Ising model makes an analytical solution out of reach.
Different approximation methods have thus been attempted, including Hamiltonian limit,~\cite{selke1981finite} free fermion approximation,~\cite{villain1981two} high temperature series expansion,~\cite{oitmaa1985high} cluster variational method~\cite{finel1986two} and others.~\cite{saqi1987iterative,selke1988annni}
Although the physics of both the small frustration regime and the energetic ground states have long been resolved, the finite temperature-strong frustration regime has not. 
Between the standard high-temperature paramagnetic phase and low-temperature modulated antiphase, a floating incommensurate (IC) phase intercalates. From field-theory, this critical phase is expected to be of the Kosterlitz-Thouless (KT) type, and thus to belong to the XY universality class.~\cite{selke1980two,kosterlitz1973ordering} Numerical validation, however, has remained elusive, as has whether the IC phase persists at large frustrations or disappears at a Lifshitz point.~\cite{selke1988annni} 
Both high and low transition temperatures ($T_\mathrm{c1}$ and $T_\mathrm{c2}$, respectively) are indeed challenging to determine in simulations~\cite{sato1999equilibrium,shirahata2001infinitesimal,chandra2007floating,rastelli2010specific,shirakura2014kosterlitz,matsubara2017domain}~(see Ref.~\onlinecite{shirakura2014kosterlitz}). 
The proposed reentrance of the IC phase around the multiphase point~\cite{finel1986two,saqi1987iterative} also remains to be confirmed. Because these features are central to our understanding of the floating IC phase in microphase formers, resolving these questions is particularly important.

As was recognized already in the mid-1980s, a transfer-matrix (TM) approach should be able to resolve all of these issues.~\cite{pesch1985transfer,beale1985finite} 
This approach indeed provides exact solutions of semi-infinite systems, which can then be extrapolated to the thermodynamic limit by finite-size scaling. 
Because both computational and memory complexity grow exponentially with system size, however, the accessible size range has long been too narrow for physical insight to emerge from such an analysis. 
Thanks to dramatic improvements in methodology, computer hardware and eigensolvers~\cite{lehoucq1998arpack} 
the TM approach has recently been applied to more complex (quasi) one-dimensional continuum-space systems, including SALR models with up to third-nearest-neighbor interaction,~\cite{hu2018clustering} and hard spheres in cylindrical confinement up to next-nearest-neighbor interaction.~\cite{godfrey2015understanding,robinson2016glasslike,hu2018correlation}  For two-dimensional lattice models with frustration, sufficiently large systems have also recently become accessible to the TM approach to determine transition temperatures on the related $J_1-J_2$ model.~\cite{jin2013phase}
In this article, we push the effective use of the TM formalism to resolve various physical ambiguities of the somewhat more complex ANNNI model. 
In particular, we determine the phase boundaries for the floating IC phase, and critically assess proposals for the Lifshitz point and the IC phase reentrance. 

\section{Transfer matrix approach}

The ANNNI model Hamiltonian for spin variables $s_i = \pm 1$ reads
\begin{equation}
\mathcal{H}_\mathrm{ANNNI} = -J_1 \sum_{\langle i, j \rangle} s_i s_j + J_2 \sum_{[ i, j]_\mathrm{axial}} s_i s_j - J_0 \sum_i s_i,
\end{equation}
where the coupling constant $J = J_1 > 0$, the frustration along the axial next-nearest-neighbor direction $\kappa = J_2/J_1>0$ and the external field $h = J_0/J$ are scaled. For $\kappa=0$, the model reduces to the standard Ising model; for the $T=0$ ground state, ferromagnetic order dominates until $\kappa < 1/2$, and the periodic antiphase (with periodicity $\langle 2 \rangle$) takes overs for $\kappa>1/2$. Note that in lattice-gas representation, this model corresponds to SALR interacting particles, 
and $h$ then plays the role of an effective chemical potential. The ANNNI model is thus clearly a minimal model for layered microphases.

The finite-temperature, finite-frustration phase behavior of semi-infinite strips is obtained by a TM approach with each layer $\mathbf{s}$ having $L$ spins $s_1, s_2, ..., s_L$. (Setting $s_{L+1} \equiv s_1$ imposes periodic boundary conditions.) Because the interaction in the ANNNI model is anisotropic, the TM can be propagated either perpendicular ($\perptm$)~\cite{pesch1985transfer} or parallel ($\paralleltm$)~\cite{beale1985finite}  to the axial next-nearest-neighbor interaction direction. 
Both matrices can be decomposed into intra-layer, $\mathbf{T}_x$, and inter-layer, $\mathbf{T}_y$, contributions,
\begin{equation}
\mathbf{T} = \mathbf{T}_x^{\frac{1}{2}} \mathbf{T}_y \mathbf{T}_x^{\frac{1}{2}}.
\end{equation}
In $\perptm$, row and column indices correspond to neighboring layer configurations $\mathbf{s}$ and $\mathbf{s}'$, 
\begin{equation} \begin{cases}
\tmat{\perp}{T}_{x} (\mathbf{s}) &= \exp [ J \sum_{i=1}^{L} (s_i s_{i+1} -\kappa s_i s_{i+2} + h s_i ) ], \\
\tmat{\perp}{T}_{y} (\mathbf{s}, \mathbf{s}') &= \exp ( J \sum_{i=1}^{L} s_i s'_i ),
\end{cases} \end{equation} 
which makes $\tmat{\perp}{T}$ a $2^L \times 2^L$ symmetric dense matrix.
In $\paralleltm$, row and column indices correspond to two subsequent layers $\{\mathbf{s}, \mathbf{s}'\}$ and $\{\mathbf{s}', \mathbf{s}''\}$, respectively, and then
\begin{equation} \begin{cases} 
\tmat{\parallel}{T}_{x} (\mathbf{s}, \mathbf{s}') &= \exp [ J \sum_{i=1}^{L} (s_i s_{i+1} + s_i s'_i + h s_i ) ], \\
\tmat{\parallel}{T}_{y} (\mathbf{s}, \mathbf{s}'') &= \exp ( -\kappa J \sum_{i=1}^{L} s_i s''_i ),
\end{cases} \end{equation} 
which makes $\tmat{\parallel}{T}$ a $4^L \times 4^L$ non-symmetric sparse matrix with $8^L$ nonzero entries.

In both cases the leading eigenvalue, $\lambda_0$, provides the free energy per spin, $f = -\log \lambda_0 / (\beta L)$, and the product of left and right leading eigenvectors, $P(\mathbf{s}) = \varphi^{-1}(\mathbf{s}) \varphi(\mathbf{s})$, provides the equilibrium probability of a layer configuration. Equilibrium configurations can thus be efficiently planted.~\cite{hu2020comment}
Thermal properties can be obtained by taking partial derivatives of $f$, e.g., the energy $u=-k_\mathrm{B} T^2 \partial (\beta f)/\partial T$ and specific heat $c=\partial u / \partial T$ per spin.
The leading correlation length can also be obtained from the spectrum gap, $\xi_1 = 1/ \log(\lambda_0 / |\lambda_1| )$, albeit only along the direction of layer propagation.
Hence, although the compactness of $\perptm$ brings larger $L$ within computational reach, the $\paralleltm$ geometry is more informative about the modulation direction, which is of greater physical interest.

Iterative eigensolvers based on matrix-vector multiplication are used to obtain first a few leading eigenvalues and eigenvectors.~\cite{Spectra2020} When only the leading eigenpairs is needed, the eigenproblem can be solved equivalently on a reduced transfer matrix,~\cite{pesch1985transfer} knowing that the original matrix is invariant to re-indexing by shifting one spin or counting spins backwards, and has $Z_2$ symmetry when $h=0$. Combining these equivalent configurations generically reduces the matrix size by a factor of $2 L$ ($4 L$ when $h=0$). As a result, $\perptm$ systems with up to $L=36$ and $\paralleltm$ systems with up to $L=16$ can be efficiently solved using $<60$GB of memory.

\section{Phase diagram for $h=0$}

\begin{figure}[t]
\includegraphics[width=\columnwidth]{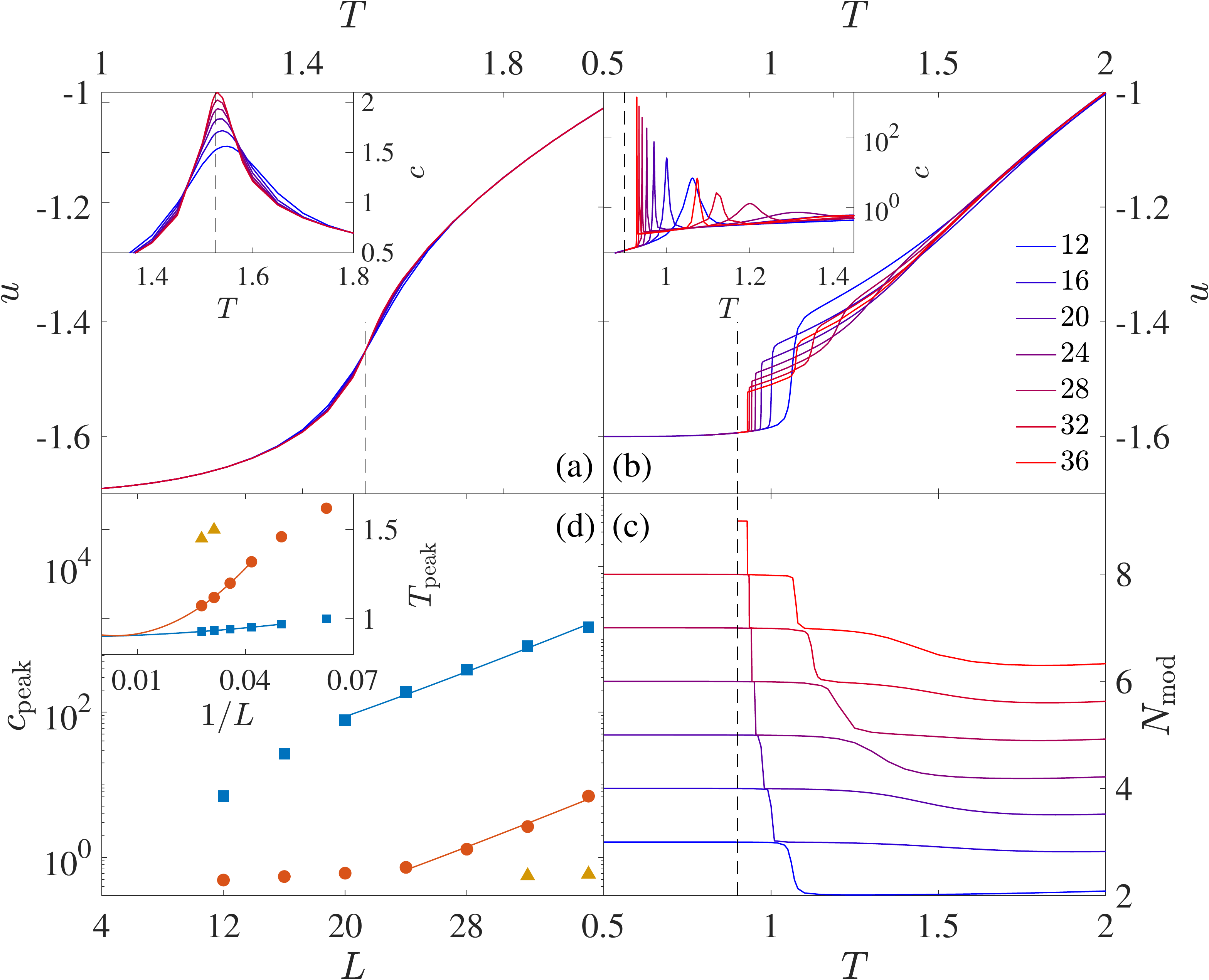}
\caption{Thermodynamic and structural observables from $\perptm$. Evolution of the energy and (inset) heat capacity per spin with temperature for (a) $\kappa=0.3$ and (b) $0.6$ with corresponding phase transition temperature estimates, $T_\mathrm{c}$ and $T_\mathrm{c2}$, respectively (dashed lines). For $\kappa=0.6$: (c) the number of modulation blocks on a layer $N_\mathrm{mod}$ decreases stepwise with $T$; (d) the finite-size scaling of the $c$ peak heights for the lowest three peaks is well described by an exponential form (see text). (Inset) Extrapolating the first peak temperature from a quadratic fit gives $T_\mathrm{c2} = 0.90$. The second peak is projected to merge with the first as $1/L\rightarrow0$. }
\label{fig:nntm}
\end{figure}

We first consider results from the $\perptm$ route (Fig.~\ref{fig:nntm}). For $\kappa <1/2$,  the energy curves for different $L$ robustly cross at well-defined critical point $T_\mathrm{c}(\kappa)$. For the Ising, $\kappa=0$, limit $u(T_\mathrm{c})$ is perfectly invariant with $L$,~\cite{ferdinand1969bounded,salas2001exact} while for $0<\kappa\lesssim1/2$, small systems, $L \lesssim 10$, exhibit a correction of at most $0.1\%$. 
From this identification of $T_\mathrm{c}(\kappa)$ we confirm that the peak of $c$ grows logarithmically with $L$, as expected for the Ising universality class. 

For $\kappa\gtrsim 1/2$ a markedly different behavior is observed.  A pronounced step in $u(T)$ gives rise to a sharp $c$ peak.~\cite{rastelli2010specific} 
At first glance, these features might suggest a simple first-order transition, in contrast to the Pokrovsky-Talapov scenario,~\cite{sato1999equilibrium} but the single step height scales as $1/L$ (not shown) and is thus projected to vanish in the thermodynamic, $1/L \rightarrow 0$, limit.
Furthermore, a second peak appears for $L \ge 24$, a third one for $L\ge32$, and it is reasonable to expect that more such peaks eventually do. This behavior is related to the stepwise change to the modulation block $N_\mathrm{mod}$ from the $\langle 2 \rangle$ antiphase (Fig.~\ref{fig:nntm}(c)). 
This change in modulation has long been considered a finite-size echo of the thermodynamic floating IC phase.~\cite{bak1982commensurate}

Although the transition temperatures identified by the step-wise steps and heat capacity peaks shift with $L$, finite-size results clearly suggest an exponential scaling, $c \sim \exp(a L)/L$ (up to $c \ge 10^3$) (Fig.~\ref{fig:nntm}(d)), and thus a first-order transition.~\cite{nightingale1982finite} At finite $L$, the modulation takes up available fractions of the system size, and as $1/L \rightarrow 0$, infinite commensurate phases are separated by infinitesimal temperature intervals. As a result the system remains critical everywhere, which is a hallmark of the floating IC phase.
The transition at $T_\mathrm{c2}$ being discontinuous (rather than critical), as was proposed in Ref.~\onlinecite{rastelli2010specific}, is thus here confirmed. 
Extrapolating the temperature of the first peak using a quadratic form further gives $T_\mathrm{c2}(\kappa=0.6) = 0.90(1)$, which is fully consistent with the most recent simulations.~\cite{matsubara2017domain} The second lowest transition is projected to merge with the first as $1/L \rightarrow 0$. The modulation wavenumber $q = N_\mathrm{mod}/L$ thus seemingly shifts from $q_0 = 1/4$ to some $q' < q_0$, resulting in an abrupt change in $q$ at $T_\mathrm{c2}$. Admittedly, the alternative scenario that $q$ could non-smoothly yet continuously change at $T_\mathrm{c2}$ cannot be excluded, but the first-order transition (discontinuity in $u$) proposal appears marginally more consistent with our results.

\begin{figure}[t]
\includegraphics[width=\columnwidth]{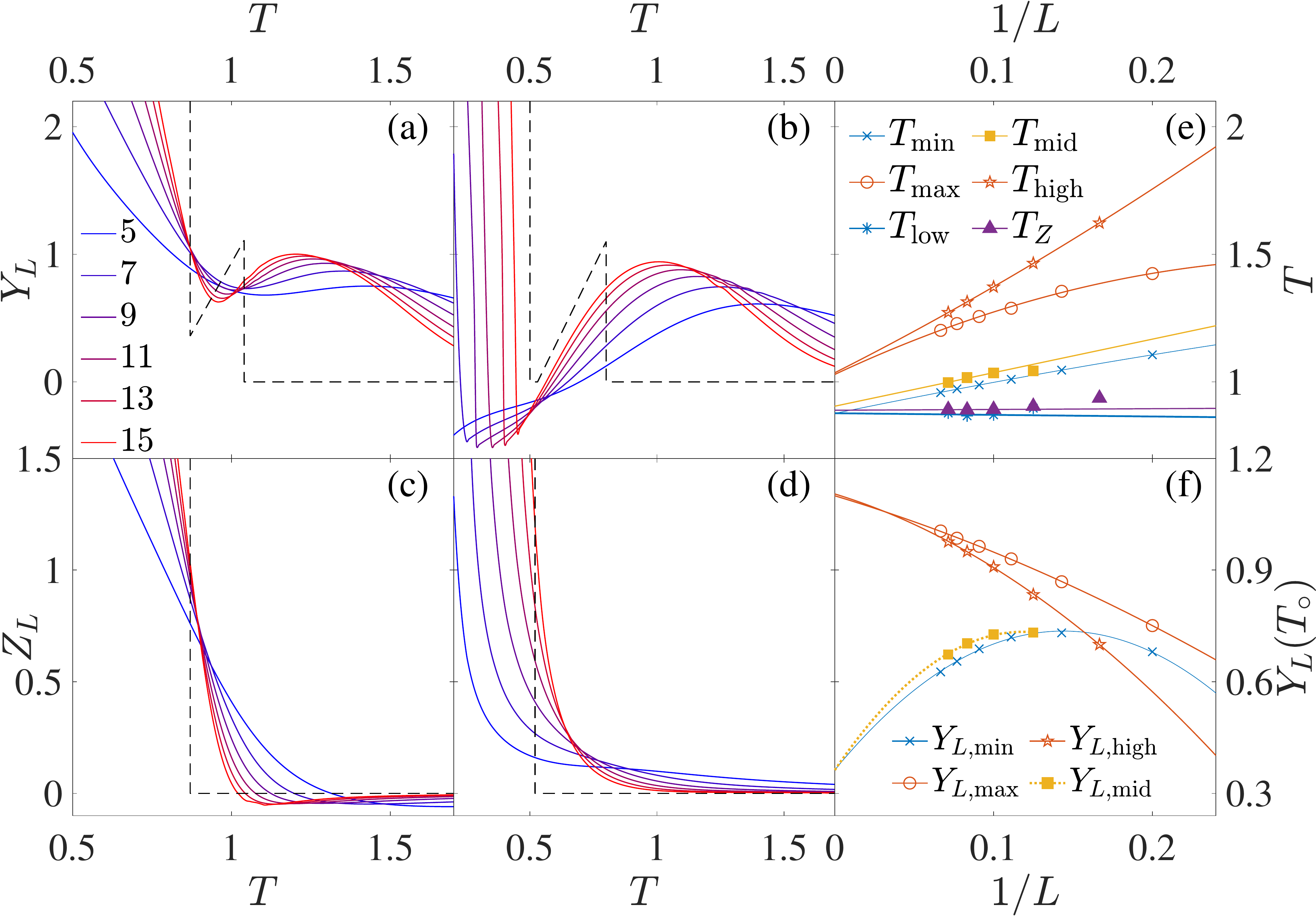}
\caption{Correlation length analysis from $\paralleltm$. Local exponents (a, b) $Y_L$ and (c, d) $Z_L$ for $\kappa=0.6$ and $0.49$, respectively. For $\kappa=0.6$: (e) the two characteristic temperatures extracted from $Y_L$ and $Z_L$ (see text for details) coincide at $T_\mathrm{c1}$ and $T_\mathrm{c2}$, respectively, as $1/L\rightarrow0$; (f) the critical exponents $Y_L$ also coincide at these temperatures. In (e) and (f), solid lines are quadratic fits (or linear fits if only three data points are available). The dotted line in $Y_{L,\mathrm{mid}}$ is purely qualitative because of the limited number of available data points. The thermodynamic behavior of $Y_L$ and $Z_L$, extracted from panels (e) and (f), are reported in (a-d) (dashed lines).}
\label{fig:nnntm}
\end{figure}

As temperature increases, the transition from the floating IC to the paramagnetic phase leaves no thermal signature, which is consistent with the KT-type universality class.
To determine its onset, we instead investigate the correlation length $\xi_1$ in $\paralleltm$. Specifically, following the finite-size analysis proposed in Ref.~\onlinecite{beale1985finite}, we define
\begin{equation}
Y_L = \frac{\log [\xi_1(L+1)] - \log [\xi_1(L-1)]}{\log (L+1) - \log (L-1)},
\end{equation}
which is the finite-$L$ (effective or local) critical exponent for $\xi_1$, i.e., $\xi_1 \sim L^{Y_L}$). Because the thermodynamic limit gives the correlation length exponent, $\lim_{L\rightarrow\infty}Y_L=\nu$, different scenarios can be discerned:
\begin{equation*} \begin{cases}
\text{in the ordered phase,}\quad & Y_L \rightarrow \infty, \\
\text{in the disordered phase,}\quad & Y_L \rightarrow 0, \\
\text{in the critical phase, or at $T_c$,}\quad & Y_L \rightarrow \mathrm{cnst}>0. \\
\end{cases} \end{equation*} 
Reference~\onlinecite{beale1985finite} also extracted the modulation wavenumber $q$ directly from the angular argument $\theta$ of the subleading eigenvalue, i.e., $q = |\theta|/ 2\pi$. This quantity brings about another local exponent $Z_L$, which characterizes the convergence to the ground state modulation, 
\begin{equation}
Z_L = -\frac{\log [\delta q(L+1)] - \log[\delta q(L-1)]}{\log (L+1) - \log (L-1)},
\end{equation}
where $\delta q = |q - q(T=0)|$. Similarly,
\begin{equation*} \begin{cases}
\text{in the commensurate phase,} & Z_L \rightarrow \infty, \\
\text{in the IC phase,} & Z_L \rightarrow 0, \\
\text{at the IC transition or disorder line,} & Z_L \rightarrow \mathrm{cnst} > 0. \\
\end{cases} \end{equation*}

Without loss of generality, we consider results for $\kappa=0.6$. Figure~\ref{fig:nnntm}(a) shows the non-monotonic evolution of $Y_L$ with $T$. Multiple crossing points ($T_\mathrm{low}$, $T_\mathrm{mid}$ and $T_\mathrm{high}$) as well as local extrema ($T_\mathrm{min}$ and $T_\mathrm{max}$) can then be identified. In addition to the crossing point in $Z_L$, namely, $T_Z$ (Fig.~\ref{fig:nnntm}(c)), the finite-$L$ scaling of these characteristic temperatures numerically determines the transition temperatures as well as the corresponding $\nu$ (Fig.~\ref{fig:nnntm}(e, f)). 
In the original approach of Ref.~\onlinecite{duxbury1984wavevector,beale1985finite}, the floating IC phase could be loosely bound by $T_Z$ and $T_\mathrm{high}$. Thanks to a vastly larger range of $L$, more characteristic temperatures can here be analyzed, thus refining numerical estimates and clarifying the underlying physics. More specifically, as $1/L\rightarrow0$, $T_\mathrm{low}, T_\mathrm{mid}$, $T_\mathrm{min}$ and $T_Z$ all coincide at $T_\mathrm{c2}$, $T_\mathrm{max}$ and $T_\mathrm{high}$ coincide at $T_\mathrm{c1}$, suggesting that $\nu$ monotonically increases with $T$ in the floating phase. The extrapolated $T_\mathrm{c2}=0.89(1)$ is fully consistent with the $\perptm$ analysis and previous MC simulations.~\cite{matsubara2017domain} By contrast, the floating-paramagnetic phase transition at $T_\mathrm{c1} = 1.04(1)$ markedly differs from prior simulation estimates, which vary from $T=1.16(1)$~\cite{shirakura2014kosterlitz} to $1.27$.~\cite{rastelli2010specific} Because finite-size corrections for this specific transition are notoriously pronounced, such discrepancy between various estimates is not surprising. The good agreement between multiple characteristic temperatures, however, clearly support our estimate.

\begin{figure}[t]
\includegraphics[width=0.95\columnwidth]{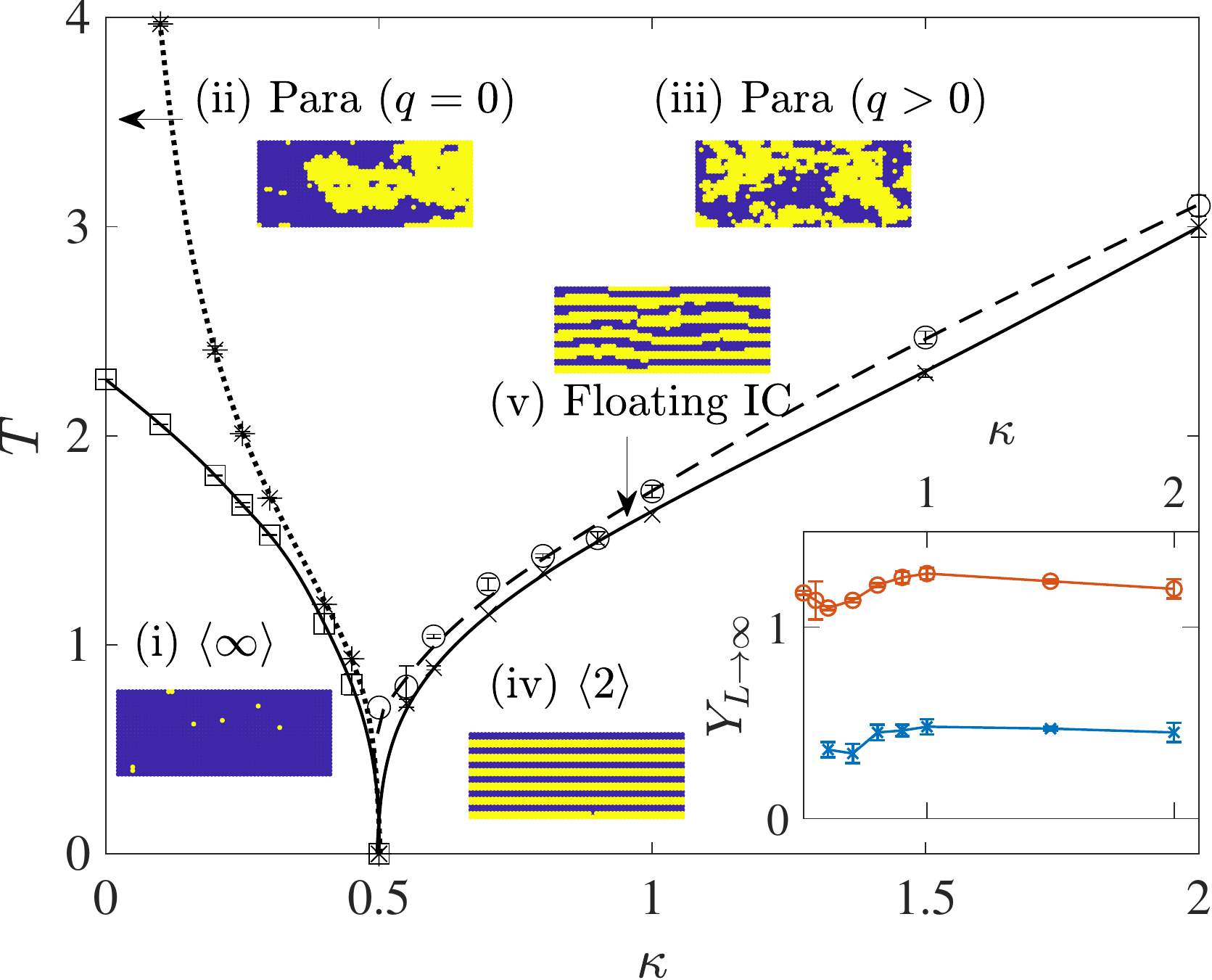}
\caption{Phase diagram for the two-dimensional ANNNI model with $h=0$. The TM approach provides phase boundaries for the ferromagnetic-paramagnetic (squares), commensurate antiphase $\langle 2 \rangle$-floating IC (crosses), and paramagnetic-IC (circles) transitions. The disorder line (asterisks) subdivides the paramagnetic phase in two regimes with  $q=0$ and $q > 0$. 
Configuration snapshots generated by planting use blue and yellow pixels to denote $+1$ and $-1$ spins, respectively. 
(Inset) Extrapolated exponents $Y_{L \rightarrow \infty}(\kappa)$ at the boundaries of the floating IC phase, $T_\mathrm{c1}$ and $T_\mathrm{c2}$. Lines are guides to the eyes.}
\label{fig:pdiag}
\end{figure}

The non-monotonic behavior of $Y_L$ persists for larger $\kappa$, but a quantitative distinction between the two transition temperatures by finite-size scaling is only feasible up to $\kappa \approx 2$. 
The extrapolation of the critical exponents $Y_L(T_\mathrm{c1})$ and $Y_L(T_\mathrm{c2})$ nevertheless remains robust, varying little with $\kappa$, which suggests that two distinct transition temperatures persist even as $\kappa$ increases (Fig.~\ref{fig:pdiag} (inset)).
This analysis strongly supports that the IC phase should survive as $\kappa \rightarrow \infty$, and goes against the finite-$\kappa$ Lifshitz point scenario.~\cite{barber1981quantum,selke1981finite}

For $\kappa$ slightly smaller than 1/2, an interesting feature emerges. For example, for $\kappa=0.49$ a narrow disordered region with $Y_L \rightarrow 0$ is squeezed between the ferromagnetic phase and the IC critical phase (Fig.~\ref{fig:nnntm}(c)). 
Knowing that the disorder line identified by the fixed point of $Z_L$ (Fig.~\ref{fig:nnntm}(d)) extends down to the multiphase point at $\kappa=1/2$ and $T=0$,~\cite{beale1985finite,finel1986two} the floating IC phase and Ising ferromagnetic phase thus never meet for $T>0$. 
This analysis confirms the reentrance of the IC critical phase in this regime, as various theoretical approximations have suggested.~\cite{finel1986two,saqi1987iterative}
The disorder line for decreasing $\kappa < 1/2$ is also found to be asymptotically tangent to $\kappa = 0$ as $T \rightarrow \infty$, instead of $\kappa = 0.25$, as was previously suggested.~\cite{finel1986two}
Combining these various observables provides a complete quantitative phase diagram for the two-dimensional ANNNI model with $h=0$ (Fig.~\ref{fig:pdiag}).

\begin{figure}[t]
\includegraphics[width=\columnwidth]{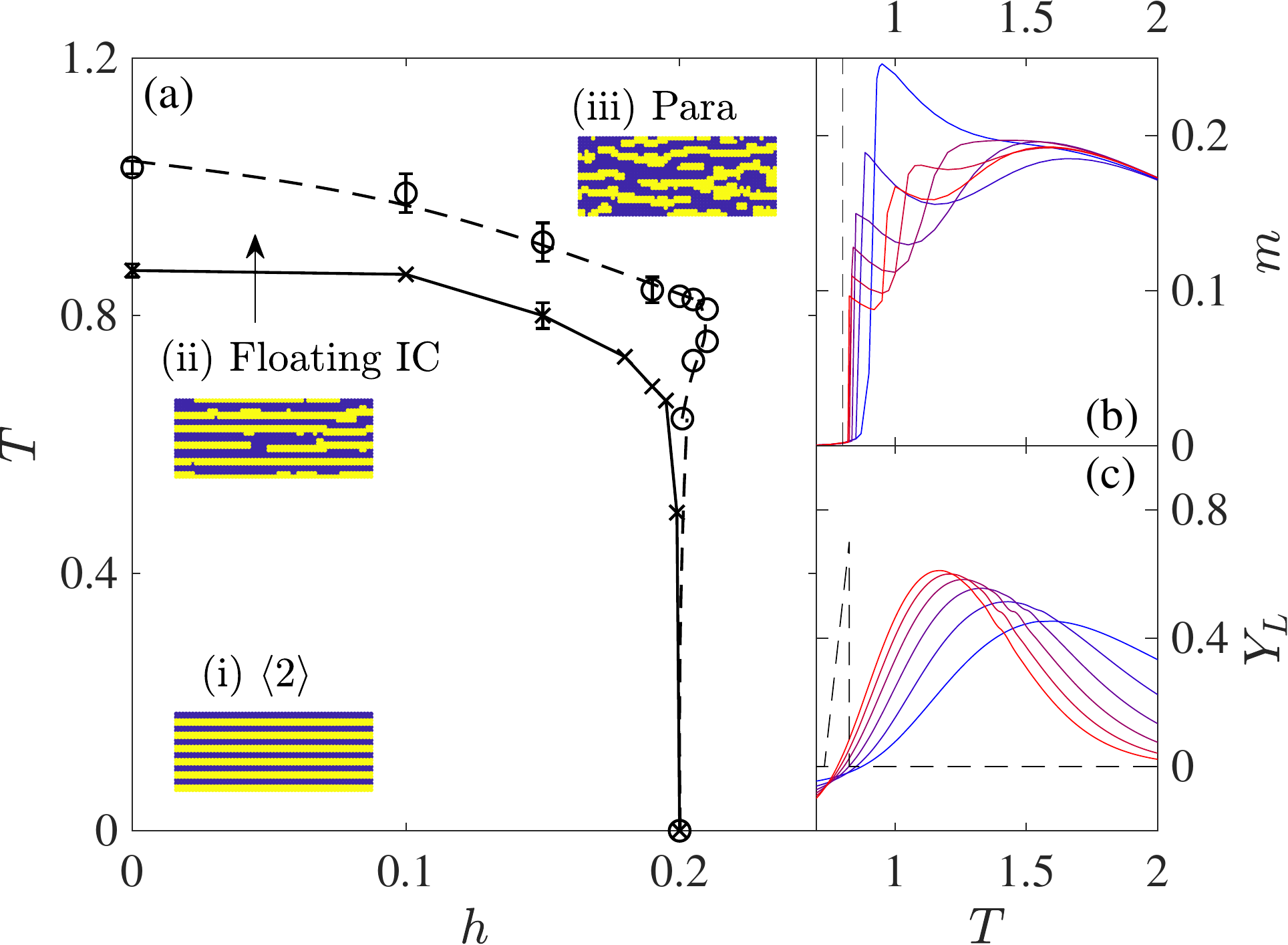}
\caption{(a) Phase diagram for the two-dimensional ANNNI model with $\kappa=0.6$ and varying $h$. Crosses denote the transition between the antiphase $\langle 2 \rangle$ and the floating IC phase; and circles that between the floating IC phase and the paramagnetic phase. Snapshots are obtained as in Fig.~\ref{fig:pdiag}. (b) Evolution of the magnetization per spin with temperature for $h=0.15$ obtained from $\perptm$ with $L=12 \ldots 32$, from blue to red. Note the clear jump in magnetization at $T_\mathrm{c2}$ (dashed line). (c) $Y_L$ for $h=0.205$ obtained from $\paralleltm$ with $L=5 \ldots 15$, from blue to red. Extrapolating characteristic temperatures (as in Fig.~\ref{fig:nnntm}) shows that the floating IC phase emerges at intermediate $T$ as $Y_L > 0$ (dashed line).}
\label{fig:mphase}
\end{figure}

\section{Phases diagram for $h>0$}

As noted above, the ANNNI  model in lattice-gas representation can be viewed as a minimal model for lamellar microphases. For this model, as in generic SALR microphase formers,~\cite{ciach2013origin} at low $T$ and small $\kappa$ a coexistence regime around $h=0$ separates the condensed ($-1$ spins dominated) and gas ($+1$ spins dominated) phases, while for $\kappa > 1/2$ lamellar microphases replace macroscopic phase separation. 
Increasing $h$ then depresses the $\langle 2 \rangle$ melting temperature down to $T \rightarrow 0$ at $h^* = 2\kappa-1$.~\cite{rujan1983rectangular} Figure~\ref{fig:mphase} shows results for $\kappa=0.6$ (for which $h^*=0.2$).
For large $h$ the ground state is a saturated paramagnetic phase and no modulation forms. 
Although this saturated paramagnetic regime exhibits spin configurations akin to those of a ferromagnetic phase, its correlation length is finite.  
At smaller $h$ -- as for $h=0$ -- a floating IC phase intercalates between the commensurate $\langle 2 \rangle$ and the paramagnetic phases. 
The $\perptm$ route confirms that the magnetization per spin, $m$, remains null in the $\langle 2\rangle$ phase and jumps (as does $u$) at the transition (Fig.~\ref{fig:mphase}(b)). 
By contrast, spin layers preferentially align with the external field in the IC phase, which leads to the magnetization stepwise increasing with temperature.
At yet higher temperatures, in the paramagnetic phase $m$ again decreases as entropy increasingly dominate.
Slightly above $h^*$ the floating IC phase reenters, in a way reminiscent of the $J_1 - J_2$ model.~\cite{guerrero2015nematic} The behavior of the local exponent $Y_L$, which crosses at $T_\mathrm{c2}$ and peaks at $T_\mathrm{c1}$ (Fig.~\ref{fig:mphase}(c)), is also similar to the reentrance in Fig.~\ref{fig:nnntm}(b). Extrapolating these special temperatures gives the phase boundaries in Fig.~\ref{fig:mphase}(a). Note that for $h \gtrsim 0.22$, the order of the extrapolated $T_\mathrm{c1}$ and $T_\mathrm{c2}$ changes and $Y_L(T_\mathrm{c1})$ is projected to vanish. The thermodynamic IC phase then terminates, even though strong finite-size echoes of it persist.

\section{Conclusion}

Using numerical transfer matrix formalism, we have resolved various ambiguities in the phase diagram of the two-dimensional ANNNI model both with and without an external field. Our results confirm the reentrance scenario for the IC phase, but also suggest that the floating IC phase persists up to $\kappa \rightarrow \infty$, and that the exponent of algebraic divergence of the correlation length remains robust. The latter two findings clearly motivate further theoretical studies. Because the TM approach provides an exact solution for semi-infinite systems, it outperforms finite-size simulations in achieving high accuracy results. Our results thus motivate reviving the TM approach for resolving equilibrium phase behavior of related frustrated models, such as the BNNNI model~\cite{oitmaa1987finite} and surfactant models.~\cite{wheeler1968phase,widom1986lattice} The system sizes now available suggests that extending the formalism to three-dimensional models is also almost within computational reach. 

\begin{acknowledgements}
We acknowledge support from the Simons Foundation (\#454937) and from the  National Science Foundation Grant No. DMR-1749374.
Data relevant to this work have been archived and can be accessed at the Duke Digital Repository.
\end{acknowledgements}

\bibliography{abbrev}

\begin{thebibliography}{49}%
\makeatletter
\providecommand \@ifxundefined [1]{%
 \@ifx{#1\undefined}
}%
\providecommand \@ifnum [1]{%
 \ifnum #1\expandafter \@firstoftwo
 \else \expandafter \@secondoftwo
 \fi
}%
\providecommand \@ifx [1]{%
 \ifx #1\expandafter \@firstoftwo
 \else \expandafter \@secondoftwo
 \fi
}%
\providecommand \natexlab [1]{#1}%
\providecommand \enquote  [1]{``#1''}%
\providecommand \bibnamefont  [1]{#1}%
\providecommand \bibfnamefont [1]{#1}%
\providecommand \citenamefont [1]{#1}%
\providecommand \href@noop [0]{\@secondoftwo}%
\providecommand \href [0]{\begingroup \@sanitize@url \@href}%
\providecommand \@href[1]{\@@startlink{#1}\@@href}%
\providecommand \@@href[1]{\endgroup#1\@@endlink}%
\providecommand \@sanitize@url [0]{\catcode `\\12\catcode `\$12\catcode
  `\&12\catcode `\#12\catcode `\^12\catcode `\_12\catcode `\%12\relax}%
\providecommand \@@startlink[1]{}%
\providecommand \@@endlink[0]{}%
\providecommand \url  [0]{\begingroup\@sanitize@url \@url }%
\providecommand \@url [1]{\endgroup\@href {#1}{\urlprefix }}%
\providecommand \urlprefix  [0]{URL }%
\providecommand \Eprint [0]{\href }%
\providecommand \doibase [0]{http://dx.doi.org/}%
\providecommand \selectlanguage [0]{\@gobble}%
\providecommand \bibinfo  [0]{\@secondoftwo}%
\providecommand \bibfield  [0]{\@secondoftwo}%
\providecommand \translation [1]{[#1]}%
\providecommand \BibitemOpen [0]{}%
\providecommand \bibitemStop [0]{}%
\providecommand \bibitemNoStop [0]{.\EOS\space}%
\providecommand \EOS [0]{\spacefactor3000\relax}%
\providecommand \BibitemShut  [1]{\csname bibitem#1\endcsname}%
\let\auto@bib@innerbib\@empty
\bibitem [{\citenamefont {Sciortino}\ \emph {et~al.}(2004)\citenamefont
  {Sciortino}, \citenamefont {Mossa}, \citenamefont {Zaccarelli},\ and\
  \citenamefont {Tartaglia}}]{sciortino2004equilibrium}%
  \BibitemOpen
  \bibfield  {author} {\bibinfo {author} {\bibfnamefont {F.}~\bibnamefont
  {Sciortino}}, \bibinfo {author} {\bibfnamefont {S.}~\bibnamefont {Mossa}},
  \bibinfo {author} {\bibfnamefont {E.}~\bibnamefont {Zaccarelli}}, \ and\
  \bibinfo {author} {\bibfnamefont {P.}~\bibnamefont {Tartaglia}},\ }\href
  {\doibase 10.1103/PhysRevLett.93.055701} {\bibfield  {journal} {\bibinfo
  {journal} {Phys. Rev. Lett.}\ }\textbf {\bibinfo {volume} {93}},\ \bibinfo
  {pages} {055701} (\bibinfo {year} {2004})}\BibitemShut {NoStop}%
\bibitem [{\citenamefont {Ciach}\ \emph {et~al.}(2013)\citenamefont {Ciach},
  \citenamefont {P{\c{e}}kalski},\ and\ \citenamefont
  {G{\'o}{\'z}d{\'z}}}]{ciach2013origin}%
  \BibitemOpen
  \bibfield  {author} {\bibinfo {author} {\bibfnamefont {A.}~\bibnamefont
  {Ciach}}, \bibinfo {author} {\bibfnamefont {J.}~\bibnamefont
  {P{\c{e}}kalski}}, \ and\ \bibinfo {author} {\bibfnamefont {W.~T.}\
  \bibnamefont {G{\'o}{\'z}d{\'z}}},\ }\href {\doibase 10.1039/C3SM50668A}
  {\bibfield  {journal} {\bibinfo  {journal} {Soft Matter}\ }\textbf {\bibinfo
  {volume} {9}},\ \bibinfo {pages} {6301} (\bibinfo {year} {2013})}\BibitemShut
  {NoStop}%
\bibitem [{\citenamefont {Zhuang}\ and\ \citenamefont
  {Charbonneau}(2016)}]{zhuang2016recent}%
  \BibitemOpen
  \bibfield  {author} {\bibinfo {author} {\bibfnamefont {Y.}~\bibnamefont
  {Zhuang}}\ and\ \bibinfo {author} {\bibfnamefont {P.}~\bibnamefont
  {Charbonneau}},\ }\href {\doibase 10.1021/acs.jpcb.6b05471} {\bibfield
  {journal} {\bibinfo  {journal} {J. Phys. Chem. B}\ }\textbf {\bibinfo
  {volume} {120}},\ \bibinfo {pages} {7775} (\bibinfo {year}
  {2016})}\BibitemShut {NoStop}%
\bibitem [{\citenamefont {Royall}(2018)}]{royall2018hunting}%
  \BibitemOpen
  \bibfield  {author} {\bibinfo {author} {\bibfnamefont {C.~P.}\ \bibnamefont
  {Royall}},\ }\href {\doibase 10.1039/C8SM00400E} {\bibfield  {journal}
  {\bibinfo  {journal} {Soft Matter}\ }\textbf {\bibinfo {volume} {14}},\
  \bibinfo {pages} {4020} (\bibinfo {year} {2018})}\BibitemShut {NoStop}%
\bibitem [{\citenamefont {Seul}\ and\ \citenamefont
  {Andelman}(1995)}]{seul1995domain}%
  \BibitemOpen
  \bibfield  {author} {\bibinfo {author} {\bibfnamefont {M.}~\bibnamefont
  {Seul}}\ and\ \bibinfo {author} {\bibfnamefont {D.}~\bibnamefont
  {Andelman}},\ }\href {\doibase 10.1126/science.267.5197.476} {\bibfield
  {journal} {\bibinfo  {journal} {Science}\ }\textbf {\bibinfo {volume}
  {267}},\ \bibinfo {pages} {476} (\bibinfo {year} {1995})}\BibitemShut
  {NoStop}%
\bibitem [{\citenamefont {Portmann}\ \emph {et~al.}(2003)\citenamefont
  {Portmann}, \citenamefont {Vaterlaus},\ and\ \citenamefont
  {Pescia}}]{portmann2003inverse}%
  \BibitemOpen
  \bibfield  {author} {\bibinfo {author} {\bibfnamefont {O.}~\bibnamefont
  {Portmann}}, \bibinfo {author} {\bibfnamefont {A.}~\bibnamefont {Vaterlaus}},
  \ and\ \bibinfo {author} {\bibfnamefont {D.}~\bibnamefont {Pescia}},\ }\href
  {\doibase 10.1038/nature01538} {\bibfield  {journal} {\bibinfo  {journal}
  {Nature}\ }\textbf {\bibinfo {volume} {422}},\ \bibinfo {pages} {701}
  (\bibinfo {year} {2003})}\BibitemShut {NoStop}%
\bibitem [{\citenamefont {Caffrey}(2009)}]{caffrey2009crystallizing}%
  \BibitemOpen
  \bibfield  {author} {\bibinfo {author} {\bibfnamefont {M.}~\bibnamefont
  {Caffrey}},\ }\href {\doibase 10.1146/annurev.biophys.050708.133655}
  {\bibfield  {journal} {\bibinfo  {journal} {Annu. Rev. Biophys.}\ }\textbf
  {\bibinfo {volume} {38}},\ \bibinfo {pages} {29} (\bibinfo {year}
  {2009})}\BibitemShut {NoStop}%
\bibitem [{\citenamefont {Fink}\ \emph {et~al.}(2019)\citenamefont {Fink},
  \citenamefont {Steiner}, \citenamefont {Szekely}, \citenamefont {Szekely},\
  and\ \citenamefont {Raviv}}]{fink2019structure}%
  \BibitemOpen
  \bibfield  {author} {\bibinfo {author} {\bibfnamefont {L.}~\bibnamefont
  {Fink}}, \bibinfo {author} {\bibfnamefont {A.}~\bibnamefont {Steiner}},
  \bibinfo {author} {\bibfnamefont {O.}~\bibnamefont {Szekely}}, \bibinfo
  {author} {\bibfnamefont {P.}~\bibnamefont {Szekely}}, \ and\ \bibinfo
  {author} {\bibfnamefont {U.}~\bibnamefont {Raviv}},\ }\href {\doibase
  10.1021/acs.langmuir.9b00778} {\bibfield  {journal} {\bibinfo  {journal}
  {Langmuir}\ }\textbf {\bibinfo {volume} {35}},\ \bibinfo {pages} {9694}
  (\bibinfo {year} {2019})}\BibitemShut {NoStop}%
\bibitem [{\citenamefont {Lecuit}\ and\ \citenamefont
  {Lenne}(2007)}]{lecuit2007cell}%
  \BibitemOpen
  \bibfield  {author} {\bibinfo {author} {\bibfnamefont {T.}~\bibnamefont
  {Lecuit}}\ and\ \bibinfo {author} {\bibfnamefont {P.-F.}\ \bibnamefont
  {Lenne}},\ }\href {\doibase 10.1038/nrm2222} {\bibfield  {journal} {\bibinfo
  {journal} {Nat. Rev. Mol. Cell Biol.}\ }\textbf {\bibinfo {volume} {8}},\
  \bibinfo {pages} {633} (\bibinfo {year} {2007})}\BibitemShut {NoStop}%
\bibitem [{\citenamefont {Heisenberg}\ and\ \citenamefont
  {Bella{\"\i}che}(2013)}]{heisenberg2013forces}%
  \BibitemOpen
  \bibfield  {author} {\bibinfo {author} {\bibfnamefont {C.-P.}\ \bibnamefont
  {Heisenberg}}\ and\ \bibinfo {author} {\bibfnamefont {Y.}~\bibnamefont
  {Bella{\"\i}che}},\ }\href {\doibase 10.1016/j.cell.2013.05.008} {\bibfield
  {journal} {\bibinfo  {journal} {Cell}\ }\textbf {\bibinfo {volume} {153}},\
  \bibinfo {pages} {948} (\bibinfo {year} {2013})}\BibitemShut {NoStop}%
\bibitem [{\citenamefont {Zhang}\ and\ \citenamefont
  {Charbonneau}(2010)}]{zhang2010monte}%
  \BibitemOpen
  \bibfield  {author} {\bibinfo {author} {\bibfnamefont {K.}~\bibnamefont
  {Zhang}}\ and\ \bibinfo {author} {\bibfnamefont {P.}~\bibnamefont
  {Charbonneau}},\ }\href {\doibase 10.1103/PhysRevLett.104.195703} {\bibfield
  {journal} {\bibinfo  {journal} {Phys. Rev. Lett.}\ }\textbf {\bibinfo
  {volume} {104}},\ \bibinfo {pages} {195703} (\bibinfo {year}
  {2010})}\BibitemShut {NoStop}%
\bibitem [{\citenamefont {Shirakura}\ \emph {et~al.}(2014)\citenamefont
  {Shirakura}, \citenamefont {Matsubara},\ and\ \citenamefont
  {Suzuki}}]{shirakura2014kosterlitz}%
  \BibitemOpen
  \bibfield  {author} {\bibinfo {author} {\bibfnamefont {T.}~\bibnamefont
  {Shirakura}}, \bibinfo {author} {\bibfnamefont {F.}~\bibnamefont
  {Matsubara}}, \ and\ \bibinfo {author} {\bibfnamefont {N.}~\bibnamefont
  {Suzuki}},\ }\href {\doibase 10.1103/PhysRevB.90.144410} {\bibfield
  {journal} {\bibinfo  {journal} {Phys. Rev. B}\ }\textbf {\bibinfo {volume}
  {90}},\ \bibinfo {pages} {144410} (\bibinfo {year} {2014})}\BibitemShut
  {NoStop}%
\bibitem [{\citenamefont {Zhuang}\ \emph {et~al.}(2016)\citenamefont {Zhuang},
  \citenamefont {Zhang},\ and\ \citenamefont
  {Charbonneau}}]{zhuang2016equilibriumPRL}%
  \BibitemOpen
  \bibfield  {author} {\bibinfo {author} {\bibfnamefont {Y.}~\bibnamefont
  {Zhuang}}, \bibinfo {author} {\bibfnamefont {K.}~\bibnamefont {Zhang}}, \
  and\ \bibinfo {author} {\bibfnamefont {P.}~\bibnamefont {Charbonneau}},\
  }\href {\doibase 10.1103/PhysRevLett.116.098301} {\bibfield  {journal}
  {\bibinfo  {journal} {Phys. Rev. Lett.}\ }\textbf {\bibinfo {volume} {116}},\
  \bibinfo {pages} {098301} (\bibinfo {year} {2016})}\BibitemShut {NoStop}%
\bibitem [{\citenamefont {Lei}\ \emph {et~al.}(2019)\citenamefont {Lei},
  \citenamefont {Krauth},\ and\ \citenamefont {Maggs}}]{lei2019event}%
  \BibitemOpen
  \bibfield  {author} {\bibinfo {author} {\bibfnamefont {Z.}~\bibnamefont
  {Lei}}, \bibinfo {author} {\bibfnamefont {W.}~\bibnamefont {Krauth}}, \ and\
  \bibinfo {author} {\bibfnamefont {A.~C.}\ \bibnamefont {Maggs}},\ }\href
  {\doibase 10.1103/PhysRevE.99.043301} {\bibfield  {journal} {\bibinfo
  {journal} {Phys. Rev. E}\ }\textbf {\bibinfo {volume} {99}},\ \bibinfo
  {pages} {043301} (\bibinfo {year} {2019})}\BibitemShut {NoStop}%
\bibitem [{\citenamefont {Fisher}\ and\ \citenamefont
  {Selke}(1980)}]{fisher1980infinitely}%
  \BibitemOpen
  \bibfield  {author} {\bibinfo {author} {\bibfnamefont {M.~E.}\ \bibnamefont
  {Fisher}}\ and\ \bibinfo {author} {\bibfnamefont {W.}~\bibnamefont {Selke}},\
  }\href {\doibase 10.1103/PhysRevLett.44.1502} {\bibfield  {journal} {\bibinfo
   {journal} {Phys. Rev. Lett.}\ }\textbf {\bibinfo {volume} {44}},\ \bibinfo
  {pages} {1502} (\bibinfo {year} {1980})}\BibitemShut {NoStop}%
\bibitem [{\citenamefont {Selke}(1981)}]{selke1981finite}%
  \BibitemOpen
  \bibfield  {author} {\bibinfo {author} {\bibfnamefont {W.}~\bibnamefont
  {Selke}},\ }\href {\doibase 10.1007/BF01292801} {\bibfield  {journal}
  {\bibinfo  {journal} {Z. Phys. B: Condens. Matter}\ }\textbf {\bibinfo
  {volume} {43}},\ \bibinfo {pages} {335} (\bibinfo {year} {1981})}\BibitemShut
  {NoStop}%
\bibitem [{\citenamefont {Villain}\ and\ \citenamefont
  {Bak}(1981)}]{villain1981two}%
  \BibitemOpen
  \bibfield  {author} {\bibinfo {author} {\bibfnamefont {J.}~\bibnamefont
  {Villain}}\ and\ \bibinfo {author} {\bibfnamefont {P.}~\bibnamefont {Bak}},\
  }\href {\doibase 10.1051/jphys:01981004205065700} {\bibfield  {journal}
  {\bibinfo  {journal} {J. Phys. (Paris)}\ }\textbf {\bibinfo {volume} {42}},\
  \bibinfo {pages} {657} (\bibinfo {year} {1981})}\BibitemShut {NoStop}%
\bibitem [{\citenamefont {Oitmaa}(1985)}]{oitmaa1985high}%
  \BibitemOpen
  \bibfield  {author} {\bibinfo {author} {\bibfnamefont {J.}~\bibnamefont
  {Oitmaa}},\ }\href {\doibase 10.1088/0305-4470/18/2/026} {\bibfield
  {journal} {\bibinfo  {journal} {J. Phys. A}\ }\textbf {\bibinfo {volume}
  {18}},\ \bibinfo {pages} {365} (\bibinfo {year} {1985})}\BibitemShut
  {NoStop}%
\bibitem [{\citenamefont {Finel}\ and\ \citenamefont
  {de~Fontaine}(1986)}]{finel1986two}%
  \BibitemOpen
  \bibfield  {author} {\bibinfo {author} {\bibfnamefont {A.}~\bibnamefont
  {Finel}}\ and\ \bibinfo {author} {\bibfnamefont {D.}~\bibnamefont
  {de~Fontaine}},\ }\href {\doibase 10.1007/BF01020657} {\bibfield  {journal}
  {\bibinfo  {journal} {J. Stat. Phys.}\ }\textbf {\bibinfo {volume} {43}},\
  \bibinfo {pages} {645} (\bibinfo {year} {1986})}\BibitemShut {NoStop}%
\bibitem [{\citenamefont {Saqi}\ and\ \citenamefont
  {McKenzie}(1987)}]{saqi1987iterative}%
  \BibitemOpen
  \bibfield  {author} {\bibinfo {author} {\bibfnamefont {M.~A.~S.}\
  \bibnamefont {Saqi}}\ and\ \bibinfo {author} {\bibfnamefont {D.~S.}\
  \bibnamefont {McKenzie}},\ }\href {\doibase 10.1088/0305-4470/20/2/032}
  {\bibfield  {journal} {\bibinfo  {journal} {J. Phys. A}\ }\textbf {\bibinfo
  {volume} {20}},\ \bibinfo {pages} {471} (\bibinfo {year} {1987})}\BibitemShut
  {NoStop}%
\bibitem [{\citenamefont {Selke}(1988)}]{selke1988annni}%
  \BibitemOpen
  \bibfield  {author} {\bibinfo {author} {\bibfnamefont {W.}~\bibnamefont
  {Selke}},\ }\href {\doibase 10.1016/0370-1573(88)90140-8} {\bibfield
  {journal} {\bibinfo  {journal} {Phys. Rep.}\ }\textbf {\bibinfo {volume}
  {170}},\ \bibinfo {pages} {213} (\bibinfo {year} {1988})}\BibitemShut
  {NoStop}%
\bibitem [{\citenamefont {Selke}\ and\ \citenamefont
  {Fisher}(1980)}]{selke1980two}%
  \BibitemOpen
  \bibfield  {author} {\bibinfo {author} {\bibfnamefont {W.}~\bibnamefont
  {Selke}}\ and\ \bibinfo {author} {\bibfnamefont {M.~E.}\ \bibnamefont
  {Fisher}},\ }\href {\doibase 10.1007/BF01295073} {\bibfield  {journal}
  {\bibinfo  {journal} {Z. Phys. B: Condens. Matter}\ }\textbf {\bibinfo
  {volume} {40}},\ \bibinfo {pages} {71} (\bibinfo {year} {1980})}\BibitemShut
  {NoStop}%
\bibitem [{\citenamefont {Kosterlitz}\ and\ \citenamefont
  {Thouless}(1973)}]{kosterlitz1973ordering}%
  \BibitemOpen
  \bibfield  {author} {\bibinfo {author} {\bibfnamefont {J.~M.}\ \bibnamefont
  {Kosterlitz}}\ and\ \bibinfo {author} {\bibfnamefont {D.~J.}\ \bibnamefont
  {Thouless}},\ }\href {\doibase 10.1088/0022-3719/6/7/010} {\bibfield
  {journal} {\bibinfo  {journal} {J. Phys. C}\ }\textbf {\bibinfo {volume}
  {6}},\ \bibinfo {pages} {1181} (\bibinfo {year} {1973})}\BibitemShut
  {NoStop}%
\bibitem [{\citenamefont {Sato}\ and\ \citenamefont
  {Matsubara}(1999)}]{sato1999equilibrium}%
  \BibitemOpen
  \bibfield  {author} {\bibinfo {author} {\bibfnamefont {A.}~\bibnamefont
  {Sato}}\ and\ \bibinfo {author} {\bibfnamefont {F.}~\bibnamefont
  {Matsubara}},\ }\href {\doibase 10.1103/PhysRevB.60.10316} {\bibfield
  {journal} {\bibinfo  {journal} {Phys. Rev. B}\ }\textbf {\bibinfo {volume}
  {60}},\ \bibinfo {pages} {10316} (\bibinfo {year} {1999})}\BibitemShut
  {NoStop}%
\bibitem [{\citenamefont {Shirahata}\ and\ \citenamefont
  {Nakamura}(2001)}]{shirahata2001infinitesimal}%
  \BibitemOpen
  \bibfield  {author} {\bibinfo {author} {\bibfnamefont {T.}~\bibnamefont
  {Shirahata}}\ and\ \bibinfo {author} {\bibfnamefont {T.}~\bibnamefont
  {Nakamura}},\ }\href {\doibase 10.1103/PhysRevB.65.024402} {\bibfield
  {journal} {\bibinfo  {journal} {Phys. Rev. B}\ }\textbf {\bibinfo {volume}
  {65}},\ \bibinfo {pages} {024402} (\bibinfo {year} {2001})}\BibitemShut
  {NoStop}%
\bibitem [{\citenamefont {Chandra}\ and\ \citenamefont
  {Dasgupta}(2007)}]{chandra2007floating}%
  \BibitemOpen
  \bibfield  {author} {\bibinfo {author} {\bibfnamefont {A.~K.}\ \bibnamefont
  {Chandra}}\ and\ \bibinfo {author} {\bibfnamefont {S.}~\bibnamefont
  {Dasgupta}},\ }\href {\doibase 10.1088/1751-8113/40/24/001} {\bibfield
  {journal} {\bibinfo  {journal} {J. Phys. A}\ }\textbf {\bibinfo {volume}
  {40}},\ \bibinfo {pages} {6251} (\bibinfo {year} {2007})}\BibitemShut
  {NoStop}%
\bibitem [{\citenamefont {Rastelli}\ \emph {et~al.}(2010)\citenamefont
  {Rastelli}, \citenamefont {Regina},\ and\ \citenamefont
  {Tassi}}]{rastelli2010specific}%
  \BibitemOpen
  \bibfield  {author} {\bibinfo {author} {\bibfnamefont {E.}~\bibnamefont
  {Rastelli}}, \bibinfo {author} {\bibfnamefont {S.}~\bibnamefont {Regina}}, \
  and\ \bibinfo {author} {\bibfnamefont {A.}~\bibnamefont {Tassi}},\ }\href
  {\doibase 10.1103/PhysRevB.81.094425} {\bibfield  {journal} {\bibinfo
  {journal} {Phys. Rev. B}\ }\textbf {\bibinfo {volume} {81}},\ \bibinfo
  {pages} {094425} (\bibinfo {year} {2010})}\BibitemShut {NoStop}%
\bibitem [{\citenamefont {Matsubara}\ \emph {et~al.}(2017)\citenamefont
  {Matsubara}, \citenamefont {Shirakura},\ and\ \citenamefont
  {Suzuki}}]{matsubara2017domain}%
  \BibitemOpen
  \bibfield  {author} {\bibinfo {author} {\bibfnamefont {F.}~\bibnamefont
  {Matsubara}}, \bibinfo {author} {\bibfnamefont {T.}~\bibnamefont
  {Shirakura}}, \ and\ \bibinfo {author} {\bibfnamefont {N.}~\bibnamefont
  {Suzuki}},\ }\href {\doibase 10.1103/PhysRevB.95.174409} {\bibfield
  {journal} {\bibinfo  {journal} {Phys. Rev. B}\ }\textbf {\bibinfo {volume}
  {95}},\ \bibinfo {pages} {174409} (\bibinfo {year} {2017})}\BibitemShut
  {NoStop}%
\bibitem [{\citenamefont {Pesch}\ and\ \citenamefont
  {Kroemer}(1985)}]{pesch1985transfer}%
  \BibitemOpen
  \bibfield  {author} {\bibinfo {author} {\bibfnamefont {W.}~\bibnamefont
  {Pesch}}\ and\ \bibinfo {author} {\bibfnamefont {J.}~\bibnamefont
  {Kroemer}},\ }\href {\doibase 10.1007/BF01307437} {\bibfield  {journal}
  {\bibinfo  {journal} {Z. Phys. B: Condens. Matter}\ }\textbf {\bibinfo
  {volume} {59}},\ \bibinfo {pages} {317} (\bibinfo {year} {1985})}\BibitemShut
  {NoStop}%
\bibitem [{\citenamefont {Beale}\ \emph {et~al.}(1985)\citenamefont {Beale},
  \citenamefont {Duxbury},\ and\ \citenamefont {Yeomans}}]{beale1985finite}%
  \BibitemOpen
  \bibfield  {author} {\bibinfo {author} {\bibfnamefont {P.~D.}\ \bibnamefont
  {Beale}}, \bibinfo {author} {\bibfnamefont {P.~M.}\ \bibnamefont {Duxbury}},
  \ and\ \bibinfo {author} {\bibfnamefont {J.}~\bibnamefont {Yeomans}},\ }\href
  {\doibase 10.1103/PhysRevB.31.7166} {\bibfield  {journal} {\bibinfo
  {journal} {Phys. Rev. B}\ }\textbf {\bibinfo {volume} {31}},\ \bibinfo
  {pages} {7166} (\bibinfo {year} {1985})}\BibitemShut {NoStop}%
\bibitem [{\citenamefont {Lehoucq}\ \emph {et~al.}(1998)\citenamefont
  {Lehoucq}, \citenamefont {Sorensen},\ and\ \citenamefont
  {Yang}}]{lehoucq1998arpack}%
  \BibitemOpen
  \bibfield  {author} {\bibinfo {author} {\bibfnamefont {R.~B.}\ \bibnamefont
  {Lehoucq}}, \bibinfo {author} {\bibfnamefont {D.~C.}\ \bibnamefont
  {Sorensen}}, \ and\ \bibinfo {author} {\bibfnamefont {C.}~\bibnamefont
  {Yang}},\ }\href {\doibase 10.1137/1.9780898719628} {\emph {\bibinfo {title}
  {ARPACK users' guide: solution of large-scale eigenvalue problems with
  implicitly restarted Arnoldi methods}}}\ (\bibinfo  {publisher} {SIAM},\
  \bibinfo {year} {1998})\BibitemShut {NoStop}%
\bibitem [{\citenamefont {Hu}\ and\ \citenamefont
  {Charbonneau}(2018)}]{hu2018clustering}%
  \BibitemOpen
  \bibfield  {author} {\bibinfo {author} {\bibfnamefont {Y.}~\bibnamefont
  {Hu}}\ and\ \bibinfo {author} {\bibfnamefont {P.}~\bibnamefont
  {Charbonneau}},\ }\href {\doibase 10.1039/C8SM00315G} {\bibfield  {journal}
  {\bibinfo  {journal} {Soft Matter}\ }\textbf {\bibinfo {volume} {14}},\
  \bibinfo {pages} {4101} (\bibinfo {year} {2018})}\BibitemShut {NoStop}%
\bibitem [{\citenamefont {Godfrey}\ and\ \citenamefont
  {Moore}(2015)}]{godfrey2015understanding}%
  \BibitemOpen
  \bibfield  {author} {\bibinfo {author} {\bibfnamefont {M.~J.}\ \bibnamefont
  {Godfrey}}\ and\ \bibinfo {author} {\bibfnamefont {M.~A.}\ \bibnamefont
  {Moore}},\ }\href {\doibase 10.1103/PhysRevE.91.022120} {\bibfield  {journal}
  {\bibinfo  {journal} {Phys. Rev. E}\ }\textbf {\bibinfo {volume} {91}},\
  \bibinfo {pages} {022120} (\bibinfo {year} {2015})}\BibitemShut {NoStop}%
\bibitem [{\citenamefont {Robinson}\ \emph {et~al.}(2016)\citenamefont
  {Robinson}, \citenamefont {Godfrey},\ and\ \citenamefont
  {Moore}}]{robinson2016glasslike}%
  \BibitemOpen
  \bibfield  {author} {\bibinfo {author} {\bibfnamefont {J.~F.}\ \bibnamefont
  {Robinson}}, \bibinfo {author} {\bibfnamefont {M.~J.}\ \bibnamefont
  {Godfrey}}, \ and\ \bibinfo {author} {\bibfnamefont {M.~A.}\ \bibnamefont
  {Moore}},\ }\href {\doibase 10.1103/PhysRevE.93.032101} {\bibfield  {journal}
  {\bibinfo  {journal} {Phys. Rev. E}\ }\textbf {\bibinfo {volume} {93}},\
  \bibinfo {pages} {032101} (\bibinfo {year} {2016})}\BibitemShut {NoStop}%
\bibitem [{\citenamefont {Hu}\ \emph {et~al.}(2018)\citenamefont {Hu},
  \citenamefont {Fu},\ and\ \citenamefont {Charbonneau}}]{hu2018correlation}%
  \BibitemOpen
  \bibfield  {author} {\bibinfo {author} {\bibfnamefont {Y.}~\bibnamefont
  {Hu}}, \bibinfo {author} {\bibfnamefont {L.}~\bibnamefont {Fu}}, \ and\
  \bibinfo {author} {\bibfnamefont {P.}~\bibnamefont {Charbonneau}},\ }\href
  {\doibase 10.1080/00268976.2018.1479543} {\bibfield  {journal} {\bibinfo
  {journal} {Mol. Phys.}\ }\textbf {\bibinfo {volume} {116}},\ \bibinfo {pages}
  {3345} (\bibinfo {year} {2018})}\BibitemShut {NoStop}%
\bibitem [{\citenamefont {Jin}\ \emph {et~al.}(2013)\citenamefont {Jin},
  \citenamefont {Sen}, \citenamefont {Guo},\ and\ \citenamefont
  {Sandvik}}]{jin2013phase}%
  \BibitemOpen
  \bibfield  {author} {\bibinfo {author} {\bibfnamefont {S.}~\bibnamefont
  {Jin}}, \bibinfo {author} {\bibfnamefont {A.}~\bibnamefont {Sen}}, \bibinfo
  {author} {\bibfnamefont {W.}~\bibnamefont {Guo}}, \ and\ \bibinfo {author}
  {\bibfnamefont {A.~W.}\ \bibnamefont {Sandvik}},\ }\href {\doibase
  10.1103/PhysRevB.87.144406} {\bibfield  {journal} {\bibinfo  {journal} {Phys.
  Rev. B}\ }\textbf {\bibinfo {volume} {87}},\ \bibinfo {pages} {144406}
  (\bibinfo {year} {2013})}\BibitemShut {NoStop}%
\bibitem [{\citenamefont {Hu}\ and\ \citenamefont
  {Charbonneau}(2020)}]{hu2020comment}%
  \BibitemOpen
  \bibfield  {author} {\bibinfo {author} {\bibfnamefont {Y.}~\bibnamefont
  {Hu}}\ and\ \bibinfo {author} {\bibfnamefont {P.}~\bibnamefont
  {Charbonneau}},\ }\href@noop {} {\bibfield  {journal} {\bibinfo  {journal}
  {arXiv preprint}\ } (\bibinfo {year} {2020})},\ \Eprint
  {http://arxiv.org/abs/2009.11194} {arXiv:2009.11194} \BibitemShut {NoStop}%
\bibitem [{\citenamefont {Qiu}(2020)}]{Spectra2020}%
  \BibitemOpen
  \bibfield  {author} {\bibinfo {author} {\bibfnamefont {Y.}~\bibnamefont
  {Qiu}},\ }\href {https://spectralib.org/doc/index.html} {\enquote {\bibinfo
  {title} {Spectralib (sparse eigenvalue computation toolkit as a redesigned
  {ARPACK}~\cite{lehoucq1998arpack})},}\ } (\bibinfo {year} {2020})\BibitemShut
  {NoStop}%
\bibitem [{\citenamefont {Ferdinand}\ and\ \citenamefont
  {Fisher}(1969)}]{ferdinand1969bounded}%
  \BibitemOpen
  \bibfield  {author} {\bibinfo {author} {\bibfnamefont {A.~E.}\ \bibnamefont
  {Ferdinand}}\ and\ \bibinfo {author} {\bibfnamefont {M.~E.}\ \bibnamefont
  {Fisher}},\ }\href {\doibase 10.1103/PhysRev.185.832} {\bibfield  {journal}
  {\bibinfo  {journal} {Phys. Rev.}\ }\textbf {\bibinfo {volume} {185}},\
  \bibinfo {pages} {832} (\bibinfo {year} {1969})}\BibitemShut {NoStop}%
\bibitem [{\citenamefont {Salas}(2001)}]{salas2001exact}%
  \BibitemOpen
  \bibfield  {author} {\bibinfo {author} {\bibfnamefont {J.}~\bibnamefont
  {Salas}},\ }\href {\doibase 10.1088/0305-4470/34/7/307} {\bibfield  {journal}
  {\bibinfo  {journal} {J. Phys. A}\ }\textbf {\bibinfo {volume} {34}},\
  \bibinfo {pages} {1311} (\bibinfo {year} {2001})}\BibitemShut {NoStop}%
\bibitem [{\citenamefont {Bak}(1982)}]{bak1982commensurate}%
  \BibitemOpen
  \bibfield  {author} {\bibinfo {author} {\bibfnamefont {P.}~\bibnamefont
  {Bak}},\ }\href {\doibase 10.1088/0034-4885/45/6/001} {\bibfield  {journal}
  {\bibinfo  {journal} {Rep. Prog. Phys.}\ }\textbf {\bibinfo {volume} {45}},\
  \bibinfo {pages} {587} (\bibinfo {year} {1982})}\BibitemShut {NoStop}%
\bibitem [{\citenamefont {Nightingale}(1982)}]{nightingale1982finite}%
  \BibitemOpen
  \bibfield  {author} {\bibinfo {author} {\bibfnamefont {P.}~\bibnamefont
  {Nightingale}},\ }\href {\doibase 10.1007/BFb0012543} {\bibfield  {journal}
  {\bibinfo  {journal} {J. Appl. Phys.}\ }\textbf {\bibinfo {volume} {53}},\
  \bibinfo {pages} {7927} (\bibinfo {year} {1982})}\BibitemShut {NoStop}%
\bibitem [{\citenamefont {Duxbury}\ \emph {et~al.}(1984)\citenamefont
  {Duxbury}, \citenamefont {Yeomans},\ and\ \citenamefont
  {Beale}}]{duxbury1984wavevector}%
  \BibitemOpen
  \bibfield  {author} {\bibinfo {author} {\bibfnamefont {P.~M.}\ \bibnamefont
  {Duxbury}}, \bibinfo {author} {\bibfnamefont {J.}~\bibnamefont {Yeomans}}, \
  and\ \bibinfo {author} {\bibfnamefont {P.~D.}\ \bibnamefont {Beale}},\ }\href
  {\doibase 10.1088/0305-4470/17/4/005} {\bibfield  {journal} {\bibinfo
  {journal} {J. Phys. A}\ }\textbf {\bibinfo {volume} {17}},\ \bibinfo {pages}
  {L179} (\bibinfo {year} {1984})}\BibitemShut {NoStop}%
\bibitem [{\citenamefont {Barber}\ and\ \citenamefont
  {Duxbury}(1981)}]{barber1981quantum}%
  \BibitemOpen
  \bibfield  {author} {\bibinfo {author} {\bibfnamefont {M.~N.}\ \bibnamefont
  {Barber}}\ and\ \bibinfo {author} {\bibfnamefont {P.~M.}\ \bibnamefont
  {Duxbury}},\ }\href {\doibase 10.1088/0305-4470/14/7/006} {\bibfield
  {journal} {\bibinfo  {journal} {J. Phys. A}\ }\textbf {\bibinfo {volume}
  {14}},\ \bibinfo {pages} {L251} (\bibinfo {year} {1981})}\BibitemShut
  {NoStop}%
\bibitem [{\citenamefont {Rujan}\ \emph {et~al.}(1983)\citenamefont {Rujan},
  \citenamefont {Selke},\ and\ \citenamefont {Uimin}}]{rujan1983rectangular}%
  \BibitemOpen
  \bibfield  {author} {\bibinfo {author} {\bibfnamefont {P.}~\bibnamefont
  {Rujan}}, \bibinfo {author} {\bibfnamefont {W.}~\bibnamefont {Selke}}, \ and\
  \bibinfo {author} {\bibfnamefont {G.~V.}\ \bibnamefont {Uimin}},\ }\href
  {\doibase 10.1007/BF01388543} {\bibfield  {journal} {\bibinfo  {journal} {Z.
  Phys. B: Condens. Matter}\ }\textbf {\bibinfo {volume} {53}},\ \bibinfo
  {pages} {221} (\bibinfo {year} {1983})}\BibitemShut {NoStop}%
\bibitem [{\citenamefont {Guerrero}\ \emph {et~al.}(2015)\citenamefont
  {Guerrero}, \citenamefont {Stariolo},\ and\ \citenamefont
  {Almarza}}]{guerrero2015nematic}%
  \BibitemOpen
  \bibfield  {author} {\bibinfo {author} {\bibfnamefont {A.~I.}\ \bibnamefont
  {Guerrero}}, \bibinfo {author} {\bibfnamefont {D.~A.}\ \bibnamefont
  {Stariolo}}, \ and\ \bibinfo {author} {\bibfnamefont {N.~G.}\ \bibnamefont
  {Almarza}},\ }\href {\doibase PhysRevE.91.052123} {\bibfield  {journal}
  {\bibinfo  {journal} {Phys. Rev. E}\ }\textbf {\bibinfo {volume} {91}},\
  \bibinfo {pages} {052123} (\bibinfo {year} {2015})}\BibitemShut {NoStop}%
\bibitem [{\citenamefont {Oitmaa}\ \emph {et~al.}(1987)\citenamefont {Oitmaa},
  \citenamefont {Batchelor},\ and\ \citenamefont {Barber}}]{oitmaa1987finite}%
  \BibitemOpen
  \bibfield  {author} {\bibinfo {author} {\bibfnamefont {J.}~\bibnamefont
  {Oitmaa}}, \bibinfo {author} {\bibfnamefont {M.~T.}\ \bibnamefont
  {Batchelor}}, \ and\ \bibinfo {author} {\bibfnamefont {M.~N.}\ \bibnamefont
  {Barber}},\ }\href {\doibase 10.1088/0305-4470/20/6/033} {\bibfield
  {journal} {\bibinfo  {journal} {J. Phys. A}\ }\textbf {\bibinfo {volume}
  {20}},\ \bibinfo {pages} {1507} (\bibinfo {year} {1987})}\BibitemShut
  {NoStop}%
\bibitem [{\citenamefont {Wheeler}\ and\ \citenamefont
  {Widom}(1968)}]{wheeler1968phase}%
  \BibitemOpen
  \bibfield  {author} {\bibinfo {author} {\bibfnamefont {J.~C.}\ \bibnamefont
  {Wheeler}}\ and\ \bibinfo {author} {\bibfnamefont {B.}~\bibnamefont
  {Widom}},\ }\href {\doibase 10.1021/ja01014a013} {\bibfield  {journal}
  {\bibinfo  {journal} {J. Am. Chem. Soc.}\ }\textbf {\bibinfo {volume} {90}},\
  \bibinfo {pages} {3064} (\bibinfo {year} {1968})}\BibitemShut {NoStop}%
\bibitem [{\citenamefont {Widom}(1986)}]{widom1986lattice}%
  \BibitemOpen
  \bibfield  {author} {\bibinfo {author} {\bibfnamefont {B.}~\bibnamefont
  {Widom}},\ }\href {\doibase 10.1103/PhysRevB.41.9148} {\bibfield  {journal}
  {\bibinfo  {journal} {J. Chem. Phys.}\ }\textbf {\bibinfo {volume} {84}},\
  \bibinfo {pages} {6943} (\bibinfo {year} {1986})}\BibitemShut {NoStop}%
\end{thebibliography}%

\end{document}